\documentclass[a4paper,10pt]{article}
\usepackage{latexsym, amsmath, amsfonts, amssymb, array, dcolumn, hhline, longtable, enumerate,psfrag,bm}
\usepackage[english]{babel}
\usepackage[dvips]{graphicx}
\usepackage{epsfig}
\usepackage[small,bf]{caption}
\begin{document}
%
%
%

\begin{center}
\noindent {\bf \Large{Enhancement of chemical activity in corrugated graphene}}

\vspace{0.5 cm}

\noindent
{\bf \large {Danil W. Boukhvalov$^*$ and Mikhail I. Katsnelson}}

\vspace{0.5 cm}

\noindent {\it Institute for Molecules and Materials, Radboud
University Nijmegen, 
~~~~~~~~~~~~~~~~~~~~~~~~~~~~~~~~~~Heyendaalseweg 135, 6525 AJ Nijmegen,
The Netherlands}

\noindent {\it E-mail: D.Bukhvalov@science.ru.nl,
M. Katsnelson@science.ru.nl}

\begin{abstract}
Simulation of chemical activity of corrugated graphene within 
density functional theory predicts an enhancement of its 
chemical activity if the ratio of height of the corrugation 
(ripple) to its radius is larger than 0.07. Further growth 
of the curvature of the ripples results in appearance of 
midgap states which leads to an additional 
strong increase of  chemisororption energy. 
These results open a way for tunable functionalization of graphene, 
namely, depending of curvature of the ripples one can provide both 
homogeneous (for small curvatures) and spot-like 
(for large curvatures) functionalization.
\end{abstract}
\end{center} 



\section{Introduction}

Graphene is considered as a prospective material for the electronics 
"beyond silicon"$^{1}$. Modern electronics is essentially two-dimensional, 
we use mainly just the surface of semiconductors and the bulk is, roughly 
speaking, a ballast limiting the perspectives of further miniaturization 
of electronic devices. It is not surprising therefore that the discovery 
of the one-atom-thick material with high electron mobility$^{2}$ has initiated a 
huge interest in this new field. At the same time, specific massless chiral c
haracter of charge carriers in graphene$^{3}$ leading to the "Klein tunneling"$^{4}$ 
is not favorable for some, potentially, the most interesting, applications 
such as transistors based on p-n junctions. Thus, creation of another 
two-dimensional materials based on graphene which would be conventional 
semiconductors is an important problem. Chemical functionalization of 
graphene is one of the most efficient ways to manipulate its physical properties$^{5}$. 
Two-side complete hydrogenation of graphene reversibly transforming it into graphane 
has been predicted theoretically$^{6,7}$ and realized experimentally$^{8}$. 
Graphane is a broad-gap semiconductor, with the gap value more than 5 eV$^{9}$. 
For the case of complete functionalization, replacement of hydrogen by other 
functional groups does not change the electronic structure essentially$^{5}$. 
To tune the value of the energy gap, one should focus on the one-side functionalization; 
the metal-insulator transition for the case of one-side hydrogenation has 
been demonstrated experimentally for graphene on substrate$^{8}$. The value of the
energy gap is changed also at the reduction of graphene oxide as was predicted 
theoretically$^{10}$ and confirmed experimentally$^{11}$. 
However, in both these cases strongly 
disordered semiconductors arise whereas to preserve high enough electron mobility 
one has to realize some regular structures of the adsorbates.
For the case of hydrogenation, the disorder is related, the most probably, 
with ripples on graphene$^{12}$ which can bind hydrogen$^{8}$. At the same  time, 
it is possible to create artificially regular ripple structures with 
a given shape$^{13}$. Ripples themselves can change drastically electronic 
structure of graphene resulting, e.g., in an appearance of midgap states$^{14}$. 
Here we study the effect of the ripples on chemical activity of graphene, 
using hydrogenation as an example, which will allow us to formulate 
specific recommendations how to produce graphene with a desirable type 
of functionalization manipulating by inhomogeneities of the substrate. 

\section{Computational method}

Our calculations have been carried out with the SIESTA
code$^{15}$ using the
generalized gradient approximation (GGA)$^{16}$ to DFT and
Troullier-Martins$^{17}$ pseudopotentials. We used energy mesh
cutoff of 400 Ry, and 10$\times$10$\times$4
$k$-point mesh in Monkhorst-Park scheme$^{18}$. 
Graphene with ripples is not strictly two-dimensional system. 
Chemisorption of chemical species leads to additional deviation 
from the planar geometry. Therefore we have chosen 
several $k$-points (namely, 4) also in $z$ direction. 
During the optimization, the electronic ground states
was found self-consistently by using norm-conserving
pseudopotentials to cores and a double-$\zeta$ plus polarization
basis of localized orbitals for carbon and metals.
Optimization of the bond lengths
and total energies was performed with an accuracy 0.04 eV /\AA
~and 1 meV, respectively.
This technical parameters of the computations are
the same as in our previous
works$^{5,7,10}$.The supercell in our computations 
contains 128 carbon atoms. 

To simulate the hydrogenation of rippled graphene 
the following procedure is used. First, we create some artificial ripple with 
a given height and radius, by a smooth out-of-plane distortion of a group of 
carbon atoms at the centre of the supercell. 
As a trial geometric shape of the ripples a semisphere has been chosen 
(Fig. \ref{fig1}a), which is isotropic, in a qualitative agreement 
with the results of Monte Carlo simulations$^{19}$ and experimental 
observations$^{12}$. The initial heighti ($h$) of the ripples 
varied from 0.01 to 0.3 nm, and the radius ($R$) from 0.8 to 1.5 nm.
Then, we add two hydrogen atoms 
to two neighboring carbon atoms (see Fig. \ref{fig1}a) 
situated at the top of the ripple and after 
that optimize geometric structure. To calculate the chemisorption energy one 
needs to know also total energy of the ripple without hydrogen which is, of 
course, unstable. To this aim, we make the ripple stabilized by the pair of 
hydrogen atoms smooth by decreasing out-of-plane displacements only for two 
atoms which were bonded with hydrogen and keeping all other atomic positions 
fixed. To calculate the chemisorption energy we use the expression 
E$_{chem}$ = E$_{graphene ~with ~ripple ~+ ~2H}$
 - E$_{graphene ~with ~ripple}$ - E$_{H_{2}}$ where the last 
term is the energy of hydrogen molecule. This characterizes stability 
of bonding of hydrogen with the ripple since it will desorb from the ripple 
as molecular hydrogen. At the same time, it allows us to estimate energetics 
of one-side hydrogenation by molecular hydrogen; the hydrogenation 
by hydrogen plasma proceeds without barriers$^{8}$. 

\section{Shape of the ripples}

First, we confirm that ripples on graphene can be stabilized by the 
chemisorption of functional groups$^{8}$. We have carried out calculations for 
the pair of fluorine atoms or the pair of hydroxyl groups, and the results are 
similar to what is presented here for the case of hydrogen. Without functionalization, 
the supercell relaxes to the flat state and the ripples disappear 
at the structure optimization. The functionalized ripples keep basically 
their original form except some geometric distortions in a close vicinity 
of the functional groups. This means that corrugations created initially 
by substrate and then decorated by chemisorption remain stable 
after elimination of the substrate.

Note that the ripples stabilized by chemisorption remain isotropic and 
have shapes not too far from semispherical ones (the real shapes are 
shown in Fig. \ref{hR}). The radius of optimized ripples 
varies between 0.7 to 1.0 nm and their height from 0.04 nm to 1.6 nm. 
These radii are of the same order as a spatial scale of geometric 
distortions around hydrogen impurity on pristine graphene$^{5,7}$. 
The effect of the impurity on electronic structure and, thus, on the 
chemisorption energies of next hydrogen atoms, is essentially nonlocal.  
The final optimized sizes of the ripples lie in much narrower intervals 
than the initially chosen parameters so one can hope that they are, 
indeed, intrinsic characteristics of the ripples and not artifacts 
of computational procedure. Fig. \ref{hR} shows optimized shapes of 
the ripples for three essentially different values of the geometric 
parameters (B, D, and E, according to the Figure \ref{fig2}a). 
One can see that neither radius nor height of the ripples can be 
used for their characterization. Similar to earlier works$^{14}$, 
we characterize the ripples by the ratio h/R which does 
correlate with the value of chemisorption energy, 
according to our calculations.  

\section{Energetics of chemisorption}

Further, we investigate dependence of the chemisorption energy on the curvature 
of the ripples. First, it is worthwhile to notice an important difference 
of one-side functionalization of graphene with and without ripples. Whereas in 
the case of flat graphene the optimal configuration of the pair of hydrogen 
atoms is para (1,4) configuration (Fig. \ref{fig1}c), for the case of ripples with 
large enough curvatures ortho (1,2) position (Fig. \ref{fig1}b) becomes more energetically 
favorable. This is due to the fact that curved surface of the ripple is close initially 
to atomic distortions created by the pair of hydrogen atoms in ortho position$^{7}$. 
This explains the initial decrease of the chemisorption energy for such pair of 
atoms (the part A-B in Fig. \ref{fig2}a), the energy gain in comparison with flat graphene 
is 0.5 eV for the point A. For stronger corrugations the chemisorption energy per 
pair drops to the values close to zero (the part C-D in Fig. \ref{fig2}a) and, at last, 
for further increase of the ratio h/R the chemisorption energy becomes negative 
(the part D-E-F). This means that molecular hydrogen will be decomposed and bonded 
at the ripples with large enough curvature stabilizing them. This sharp decrease of 
the total energy can be related with the formation of the midgap states at the 
corresponding ripples (midgap) as one can see in Fig. \ref{fig2}b. Hydrogen atoms destroy 
the midgap states providing the energy gain (see Fig. \ref{fig2}b). In such cases 
the chemisorption of six hydrogen atoms leads to the opening of the gap in 
electron energy spectrum which does not take place for less curvature 
of the ripples (Fig. \ref{fig2}c). 

We tried to create initially ripples with the ratio h/R uniformly distributed 
between 0.01 and 0.17. However, after optimization of the structure three 
ranges have appeared (A-B, C-D, and D-E) with breaks between them, with 
average values of h/R equal to 0.08, 0.12, and 0.16, respectively. 
Interestingly, for chemisorption of the pair of hydrogen atoms at flat graphene 
the height of out-of-plane atomic distortions is about 0.04 nm, with the radius 
of the distorted region $\approx$ 1 nm, so, h/R $\approx$
0.04$^{5}$. Thus, the regions of stable 
ripples are characterized approximately by integer numbers of this ratio. 
For h/R $<$ 0.07 the ripples disappear at the optimization of atomic positions 
and graphene with the pair of hydrogen atoms remains flat. 
The value h/R $\approx$0.12 (part C-D) is typical for fullerenes, 
therefore, the chemisorption energies for this range is close to that 
for chemisorption of the pair of hydrogens at C$_{60}$$^{20}$.  Further increase 
of the curvature (the part E-F) when the midgap states appear corresponds 
to the ripples (for our choice R $\approx$ 1 nm) where the length of C-C bonds 
reaches its maximal value 0.155 nm. Further increase of the curvature will 
lead to breaking of the chemical bonds and formation of vacancies and other 
types of defects. It was noticed earlier$^{21}$ that appearance of the midgap states 
requires so strong stresses that formation of dislocations becomes possible. 
Chemisorption of hydrogen destroys the midgap states stabilizing such ripples 
which may be an alternative to vacancy or dislocation formation. 

\section{Conclusion}

\begin{figure}[ht]
 \begin{center}
   \centering
\includegraphics[width=5.2 in]{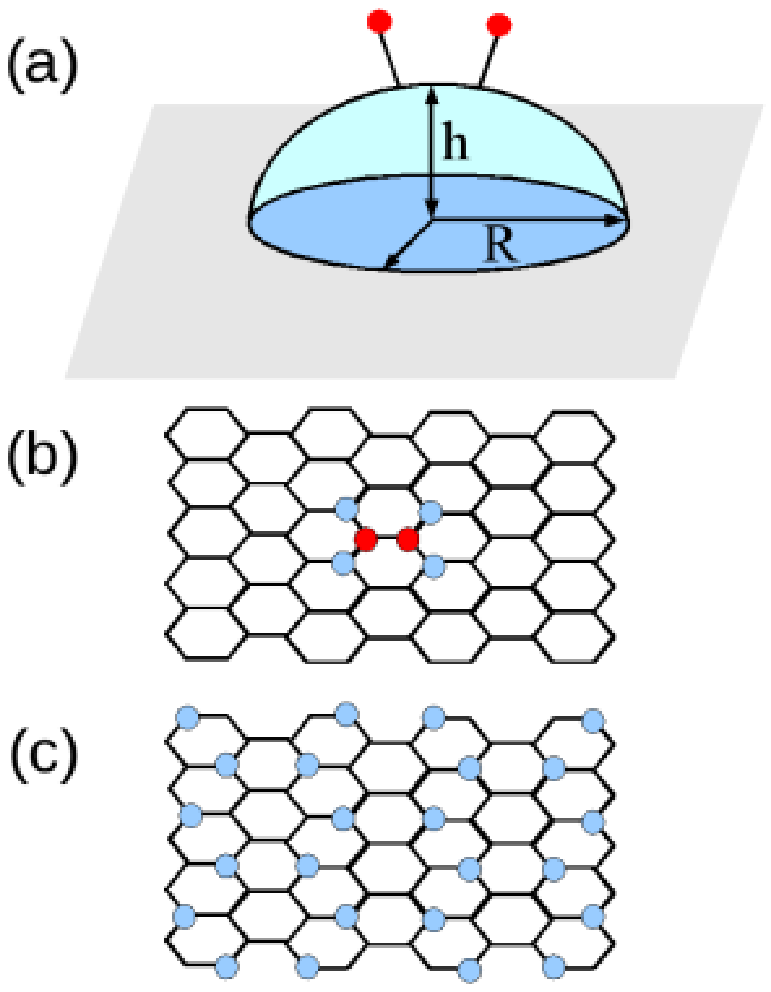}
\caption{Schemes of chemisorption of pair of hydrogen atoms 
on the top of the ripple with the height h and radius R (a), 
of step-by-step hydrogenation of the top of the ripple 
(red pair is the first step and the blue pairs is the 
second one) (b), and of complete homogeneous one-side 
functionalization of flat graphene (c).}
            \label{fig1}
 \end{center}
\end{figure}

\begin{figure}[ht]
 \begin{center}
   \centering
\rotatebox{-90}{
\includegraphics[height=5.2 in]{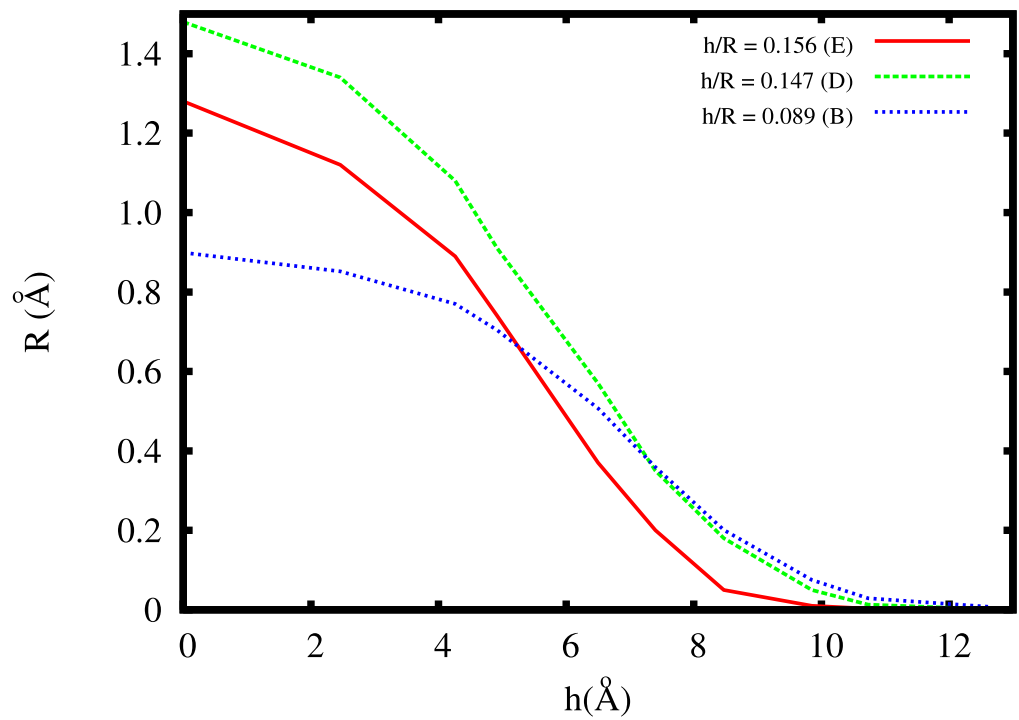}}
\caption{Optimized shapes of the ripples stabilized 
by hydrogenation, for special points 
B, D, E in Fig. \ref{fig2}a.}
            \label{hR}
 \end{center}
\end{figure}

\begin{figure}[ht]
 \begin{center}
   \centering
\includegraphics[width=5.2 in]{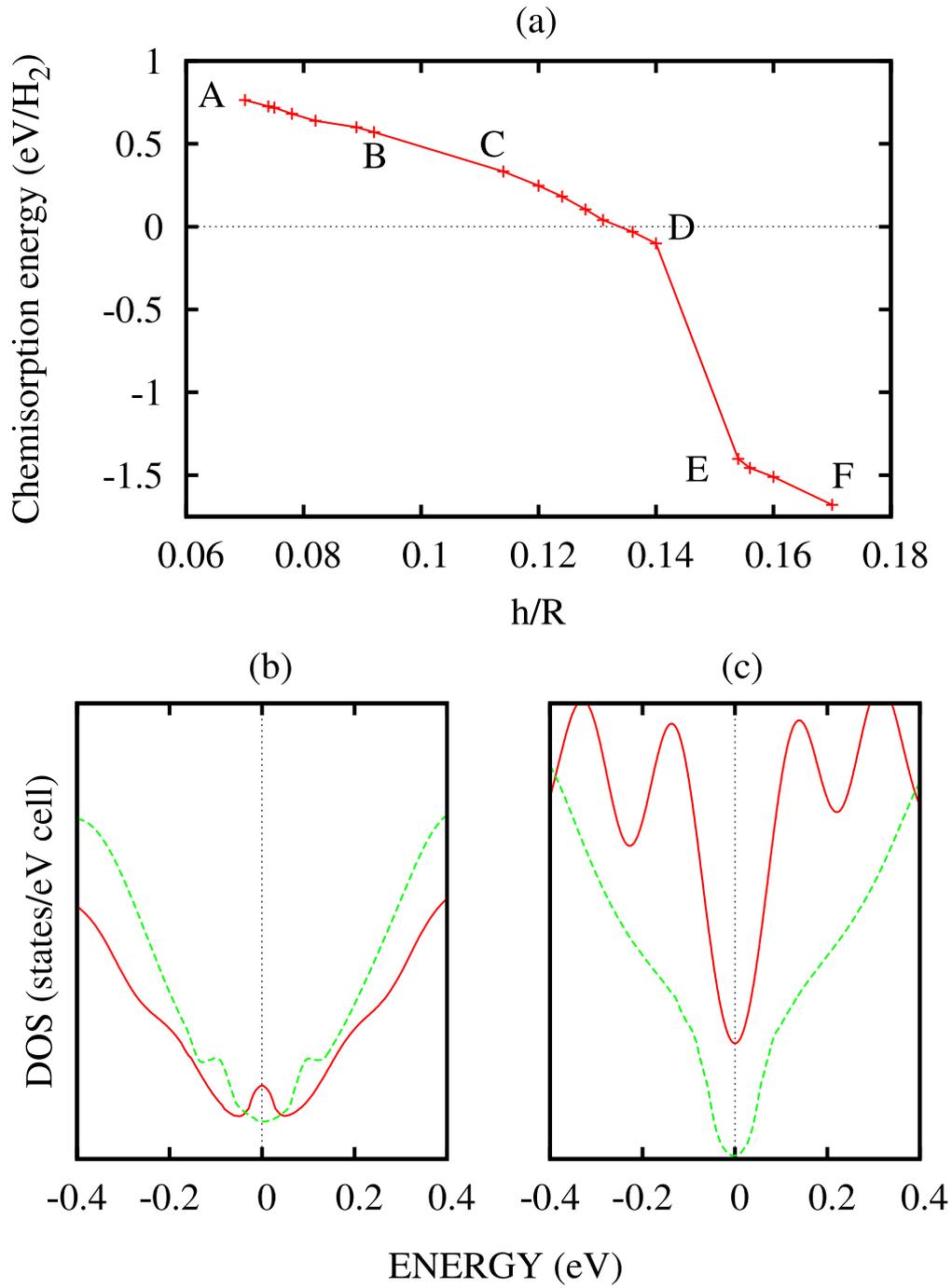}
\caption{Dependence of the chemisorption energy on curvature of 
the ripple (the radius was of order of 1 nm, more detailed data 
see in the Supporting Information) (a); 
density of states for the point E with (dashed green line) 
and without (solid red line) the pair of hydrogen atoms (b); 
density of states for six hydrogen atoms per ripple (see Fig. 1b) 
for the points D (solid red line) and E (dashed green line) (c).}
            \label{fig2}
 \end{center}
\end{figure}

To conclude, we have demonstrated that ripples in graphene effect drastically 
on its chemical activity. Functionalization can stabilize ripples with very 
strong strains, close to the breaking of carbon-carbon bonds. 
However, if original corrugations are small enough, such that typical h/R is 
smaller than 0.07, the ripples will disappear after elimination of 
the substrate even in the presence of hydrogen. In this situation 
one can hope to provide a regular structure shown in Fig. \ref{fig1}c. 

{\bf Acknowledgment} ~~
The work is financially supported by Stichting voor
Fundamenteel Onderzoek der Materie (FOM), the Netherlands.


\end{document}